\UseRawInputEncoding
\RequirePackage{fix-cm}
\documentclass[smallextended]{svjour3}       
\smartqed  
\usepackage{geometry}
\usepackage{graphicx, enumitem,chngcntr}
\usepackage{amsmath,amstext,amssymb,url}
\usepackage[justification=centering]{caption}
\usepackage{cite}
\counterwithin{figure}{section}
\numberwithin{equation}{section}
\begin{document}

\title{AEAD Modes for ZUC Family Stream Ciphers
}


\author{Hongli Li \and Yonghui Wang \and Yongbiao Ma \and Wenyi Jia \and Liang Bai 
}


\institute{Hongli Li\textsuperscript{1} \and Yonghui Wang\textsuperscript{2} \and Yongbiao Ma \and Wenyi Jia \and Liang Bai \at
              XingTang Teleommunications Technology Co. LTD., Beijing, China \\
              \email{\textsuperscript{1}hllili@mail.ustc.edu.cn, \textsuperscript{2}yhwang2001@126.com}
}

\date{Received: date / Accepted: date}

\maketitle

\begin{abstract}
In order to improve the efficiency of using ZUC primitives, we give two AEAD (Authenticated Encryption with Associated Data) modes for them, ZUC-GXM and ZUC-MUR. They are suitable for ZUC (ZUC-128) and two cases of ZUC-256. The former is a nonce-based AEAD, which is following the GCM framework. The latter is a nonce misuse-resistant one which is based on the framework of SIV variance, providing more robust applications for ZUC family stream ciphers. Both modes are designed for different using cases today, or the upcoming applications in 5G.
\end{abstract}

\section{Introduction}
\label{intro}
ZUC with 128-bit secret key is a stream cipher used in the 3GPP \cite{13}. Its confidentiality and integrity algorithms are 128-EEA3 and 128-EIA3, respectively. The ZUC-256 \cite{1} offers 256-bit security for the upcoming applications in 5G, which follows the design principles of the previous ZUC with small differece. We consider how to design modes coupling confidentiality and authenticity based on ZUC primitives, such that one can use them efficiently.

AEAD, Authenticated Encryption with Associated Data, is used to provide privacy of plaintext, and authenticity of plaintext and associated data. It's usually built on some security primitives, such as blockciphers and stream ciphers. Recently, according to different use cases of initial vector (or nonce), AEAD is divided into nonce-based (nAEAD)and nonce misuse-resistance (mrAEAD) one. Briefly, for the former, it fails catastrophically in face of nonce reuse. While for the latter, it remains the most security when the nonce is used more than once. That is, in some meaning, repeating a nonce does not affect anthenticity of the scheme and only allow an adversary to detect if the same message was already encrypted along with the same nonce before.
mrAEAD has alreaday become a design trend. Generally, it is less efficient, while it can remedy the nonce-misuse problem faced by most of the nAEAD, and so it is suitable for more robust applications. nAEAD examples are GCM \cite{2}, OCB \cite{9}, and CCM \cite{12} etc.. Examples of schemes achieving mrAEAD security notion are BTM \cite{15}, SIV \cite{8}, GCM-SIV \cite{4}, SCT \cite{5} etc.. In the CAESAR competition \cite{8}, some other modes have been well studied. AEAD based on stream ciphers is not so popular, only HS1-SIV \cite{11}, ChaCha20/Poly1305 \cite{10}, and Snow-V-AEAD \cite{7} are publicly known.

In this paper, we give two AEAD modes for ZUC, ZUC-GXM and ZUC-MUR, where ZUC-GXM is a nonce-respect one and ZUC-MUR is nonce misuse-resistant. Both of them are suitable for ZUC family, ZUC-128, and ZUC-256 with 184 or 128 bits initial vector. They are aimed for different using cases today, or the upcoming applications in 5G.

Organization of the rest of the paper is as follows. In Section 2 the preliminaries are given. In Section 3 we describe the ZUC-GXM scheme, and its security analysis. The robust AEAD scheme ZUC-MUR and its security analysis is given in Section 4. And the paper ends with conclusions in Section 5.
\section{Preliminaries and Definitions}\label{sec:2}
\subsection{Notations}\label{subsec:2.1}
We write $\mathrm{\{0,1\}^\star}$ for the set of all finite bit strings including the empty string $\varepsilon$. For a bit string
$\mathrm{X\in\{0,1\}^\star}$, $\mathrm{|X|}$ denotes its length in bits, $\mathrm{|X|_{128}= \lceil X \rceil/128}$.
For $\mathrm{X,Y\in\{0,1\}^\star}$, $\mathrm{X\|Y}$ denotes their concatenation. We write the $l$-bit zero string as 0$^l$ = 0$\|\cdots\|$0 $\in\{0,1\}^l$. For $\mathrm{X\in\{0,1\}^\star}$ with $\mathrm{|X|\geq}$ $l$, MSB$_{l}$(X) denotes the first (leftmost) $l$ bits of X. $\mathrm{Partition(X)= X[1]\|\cdots \|X[x]}$ denotes the decomposition of X into $128$-bit blocks, where $\mathrm{x=|X|_{128}}$, i.e., $\mathrm{X[1],\cdots,X[x]}$ are unique
bit strings such that $\mathrm{X[1]\|\cdots \|X[x]=X, |X[1]|=\cdots =|X[x-1]|=128}$, and $\mathrm{0 < |X[x]|\leq 128}$. If $\mathrm{|X[x]|\neq 128}$,
we pad it with zeros till $128$-bit string $\mathrm{X^\prime[x]}$, that is, $\mathrm{X^\prime[x]= X[x]\|0^{128-|X[x]|}}$. If X = $\varepsilon$, then $\mathrm{Partition(X) = X[1] = \varepsilon}$.
For a finite set $\chi$, $\mathrm{X\stackrel{\$}{\leftarrow}\chi}$ means a uniform random sampling of an element X from $\chi$. For integers $l$ and X such that X $< 2^l$, $str_l()$ is the standard $l$-bit binary representation of X. $\mathrm{\widehat{X}}$ denotes a set of strings of length $\mathrm{|X|}$.

The set of 128-bit strings, $\mathrm{\{0,1\}^{128}}$, is also regarded as $\mathrm{GF(2^{128})}$, the finite field with $\mathrm{2^{128}}$. An 128-bit string $\mathrm{a_{127},\ldots,a_1,a_0\in\{0,1\}^{128}}$ corresponds to a formal polynomial $\mathrm{a(x)= a_{127}+a_{126}x+¡­+a_1x^{126}+a_0x^{127}}$ $\mathrm{\in GF(2)[x]}$. The polynomial used is $\mathrm{p(x)=1+x+x^2+x^7+x^{128}}$ for the multiplication of two element, which is the same as in GCM.

For two binary strings $\mathrm{X = X_{1}\|\ldots\|X_m}$ and $\mathrm{Y = Y_{1}\|\ldots\|Y_m}$, the notation $\mathrm{X\oplus Y}$ denotes bitwise xor. For any string X, define $\mathrm{X\oplus \varepsilon = \varepsilon\oplus X = X}$.
\subsection{Some definitions}\label{subsec:2.2}
Some useful definitions are given in this part. Most of them come from public reference with a little difference.

\noindent\textbf{Definition 1 (random function).} Let Func(n, m) be the set of all functions $\{0,1\}^{n} \rightarrow\{0,1\}^{m}$. In addition, let Perm(n) be the set of all permutations over $\{0,1\}^{n}$. A uniform random function (URF) R having n-bit input and m-bit output is uniformly distributed over Func(n, m). It is denoted by R$\stackrel{\$}{\leftarrow}$Func(n, m).

\noindent\textbf{Definition 2 (cpa security).} Let $\mathrm{F_K: \{0,1\}^{n} \rightarrow\{0,1\}^{m}}$ and $\mathrm{G_{K^\prime}: \{0,1\}^{n} \rightarrow\{0,1\}^{m}}$ be two compatible keyed functions, with $\mathrm{K\in \widehat{K}}$ and $\mathrm{K^\prime\in \widehat{K}^\prime}$ (key spaces are not necessarily the same). Let $\mathcal{A}$ be an adversary trying to distinguish them using chosen plaintext queries.Then the advantage of $\mathcal{A}$ is defined as
\begin{equation*}
\mathrm{Adv_{F_K, G_{K^\prime}}^{cpa}(\mathcal{A})=Pr[K\stackrel{\$}{\leftarrow}\widehat{K}: \mathcal{A}^{F_K}\Longrightarrow1]-Pr[K^{\prime}\stackrel{\$}{\leftarrow}\widehat{K}^\prime: \mathcal{A}^{G_{K^{\prime}}}\Longrightarrow1]}.
\end{equation*}
\noindent\textbf{Definition 3 (pseudorandom function, prf).} The above definition can natruely be extended to $\mathrm{F_K}$ or $\mathrm{G_{K^\prime}}$ is R, where R is a URF, we have
\begin{equation*}
\mathrm{Adv_{F_K}^{prf}(\mathcal{A})=Adv_{F_K, R}^{cpa}(\mathcal{A})}.
\end{equation*}
Further, we define
\begin{equation*}
\mathrm{Adv_{F_K}^{prf}(t, q, \sigma)=max_{\mathcal{A}}Adv_{F_K}^{prf}(\mathcal{A})}.
\end{equation*}
where $\mathcal{A}$ makes at most $q$ quries, totally $\sigma$ bits encryption quries in running time at most $t$.

\noindent\textbf{Definition 4 (pseudorandom generator, prg).} The keystream generator prg advantage of KS is defined as
\begin{equation*}
\mathrm{Adv_{KS(r)}^{prg}(\mathcal{A})=Pr[K\stackrel{\$}{\leftarrow}\widehat{K}: \mathcal{A}^{KS_K(r)}\Longrightarrow1]-Pr[\mathcal{A}^{\$}\Longrightarrow1]}.
\end{equation*}
If $\mathrm{max_{\mathcal{A}}Adv_{KS(r)}^{prg}(\mathcal{A})}$ is negligible, then we say $\mathrm{KS_K(r)}$ is a prg, where r is the length of generated keystream, random oracle $\$$ has the same output length as KS, K is the secret seed in space $\mathrm{\widehat{K}}$.

While the prgs we discussed sofar only take the seed as input, many prgs used in practice take an additional input called a nonce . That is, the prg is a function $G: \mathrm{\widehat{K}\times\widehat{N}\longrightarrow\widehat{R}}$, where $\mathrm{\widehat{K}}$ and $\mathrm{\widehat{R}}$ are as before and $\mathrm{\widehat{N}}$ is called a nonce space . The nonce lets us generate multiple pseudorandom outputs from a single seed K. That is, $\mathrm{G(K, n_0)}$ is one pseudorandom output and $\mathrm{G(K, n_1)}$ for $\mathrm{n_1\neq n_0}$ is another. The nonce turns the prg into a more powerful primitive called a pseudorandom function. Secure pseudorandom functions make it possible to use the same seed to encrypt multiple messages securely. In this paper, we assume ZUC has this stronger property than as a prg.

\noindent\textbf{Definition 5 (cca).} $\mathrm{E_K^{-1}}$ and $\mathrm{G_{K^\prime}^{-1}}$ be the inversion of permutations $\mathrm{E_K}$ and $\mathrm{G_{K^\prime}}$, respectively. For a cca-adversary $\mathcal{A}$ who can makes both encryption and decryption on¡¼$\mathrm{E_K}$ and $\mathrm{E_K^{-1}}$,and on $\mathrm{G_{K^\prime}}$ and $\mathrm{G_{K^\prime}^{-1}}$£¬then we define $\mathcal{A}$'s cca-advantage as
\begin{equation*}
\mathrm{Adv_{E_K, G_{k^\prime}}^{cca}(\mathcal{A})=Pr[K\stackrel{\$}{\leftarrow}\widehat{K}: \mathcal{A}^{E_K,E_K^{-1} }\Longrightarrow1]-Pr[K^{\prime}\stackrel{\$}{\leftarrow}\widehat{K}^\prime: \mathcal{A}^{G_{K^{\prime}}, G_{K^\prime}^{-1}}\Longrightarrow1]}.
\end{equation*}
\textbf{Nonce-respected AEAD (nAEAD) security.} Nonce-respected AEAD, briefly called nAEAD. Let the encryption and decryption functions be AEnc and ADec repesctively.

A cpa-adversary $\mathcal{A}$ attacks AEAD by querying encryption AEnc. Let \newline $\mathrm{(IV_1,A_1, P_1),\ldots(IV_q,A_q, P_q)}$ be all the encryption queries made by $\mathcal{A}$, and their responces are respectively $\mathrm{(IV_1,A_1, C_1, T_1),\ldots(IV_q,A_q, C_q, T_q)}$. For an nAEAD, we assume all $\mathrm{IV_is}$ are distinct. Define random oracle $\$$, whose input is (IV, A, P), output is $\mathrm{(C, Tag)\stackrel{\$}{\leftarrow}\{0,1\}^{|P|}\times\{0,1\}^{\tau}}$, where $\tau$ is the length of the Tag.

\noindent\textbf{Definition 6 (privacy of nAEAD).} For a Nonce-respected cpa-adversary $\mathcal{A}$, we define the nonce-respect privacy as
\begin{equation*}
\mathrm{Adv_{nAEAD}^{priv}(\mathcal{A})=Pr[K\stackrel{\$}{\leftarrow}\widehat{K}: \mathcal{A}^{AEnc_K}\Longrightarrow1]-Pr[\mathcal{A}^{\$}\Longrightarrow1]}.
\end{equation*}

\noindent\textbf{IV-based vs. Nonce-based encryption.} The difference between IV-based and nonce-based encryption is that a random IV is used in the former, where as in the latter a unique nonce is provided as input in each encryption. The guarantee of nonce-based encryption is security is maintained as long as the nonce used is different each time.

\noindent\textbf{Definition 7 (authenticity).} For the cca-adversary $\mathcal{A}$'s authenticity, we define as
\begin{equation*}
\mathrm{Adv_{nAEAD}^{auth}(\mathcal{A})=Pr[K\stackrel{\$}{\leftarrow}\widehat{K}: \mathcal{A}^{AEnc_K,ADec_K}forges]},
\end{equation*}
where $\mathcal{A}$ forges if ADec returns a bit string (other than $\bot$) for a decryption query $\mathrm{(IV_{i}^\prime,A_{i}^\prime, C_{i}^\prime, T_{i}^\prime)}$ for some $\mathrm{1\leq i\leq q_v}$ such
that $\mathrm{(IV_{i}^\prime,A_{i}^\prime, C_{i}^\prime, T_{i}^\prime)\neq(IV_{j},A_{j}}$, $\mathrm{C_{j}, T_{j})}$, for all $\mathrm{1\leq j\leq q}$.

\noindent\textbf{Definition 8 (mrAEAD Security).} For an mrAEAD, a adversary $\mathcal{A}$'s mrAEAD distinguish advantage is defined as
\begin{equation*}
\mathrm{Adv^{mrAEAD}(\mathcal{A})=Pr[K\stackrel{\$}{\leftarrow}\widehat{K}: \mathcal{A}^{AEnc_K,ADec_K}\Longrightarrow1]-Pr[\mathcal{A}^{\$,\perp}\Longrightarrow1]},
\end{equation*}
where AEnc$_K$ denotes the encryption query with input (N, A, P), output ciphertext (C, T), and ADec$_K$ denotes the decryption query with input (N, A, C, T), output plaintext of ciphertext C or an invalid symbol $\perp$. $\$$ is a random string of length $\mathrm{|AEnc_K(N, A, P)|}$ and $\perp$ is an invalid symbol. The adversary is not allowed to query the same ternary twice.

Other useful definition is $\epsilon-$AXU (Almost XOR Universal) Hash Function, which is defined as follow:

Consider a keyed hash function H with a set of keys $\mathrm{\widehat{K}\subseteq\{0,1\}^k}$, Domain $\mathrm{\widehat{D}\subseteq\{0,1\}^*}$, and range $\mathrm{\{0, 1\}^{n}}$. It takes a key $\mathrm{L\in\widehat{K}}$ and $\mathrm{X\in\widehat{D}}$ as input, and returns the output $\mathrm{x=H_{L}(X)\in\{0,1\}^{n}}$. H is an $\mathrm{\epsilon-AXU}$ hash function if for any distinct $\mathrm{X, Y\in\hat{D}}$, and $\mathrm{z\in\{0, 1\}^{n}}$, it holds that $\mathrm{Pr_{L}[H_{L}(X)\oplus H_{L}(Y)=z]\leq\epsilon}$, where $\mathrm{\epsilon}$ is a negligeable.

In this paper, we use $\epsilon-$AXU hash functions GHASH, which is the same as defined in GCM.
\subsection{Brief Discription of ZUC Family}\label{subsec:2.3}
ZUC family include ZUC-128 (with 128 bits key and 128 bits initial vector) and two cases of ZUC-256, with 256 bits key and 184 or 128  bits initial vector respectively. The ZUC-256\cite{1} stream ciphers are successors of the previous ZUC-128 used in the 3GPP, which including confidentiality algorithms 128-EEA3 and integrity algorithms 128-EIA3. Both of ZUC-256 cases aim to give new stream ciphers that offering 256-bit security for the upcoming applications in 5G. And three algorithms are mainly including two phases, Initialization phase and keystreamGeneration phase. In order to be highly compatible with ZUC-128, ZUC-256 primitives differ from ZUC-128 only in the initialization phase. At each time instant, they all generate a 32-bit keystream word. More details about them please refer to \cite{1,13,14}.

Convinently, we use Z = ZUC$_l$(IV, K) to denote using initial vector  IV and secret key K to generate $l$-bit keystream Z. The procedure is as follows.

\vspace{2ex}
\textbf{Z = ZUC$_l$(IV, K) Algorithm:}
\begin{itemize}[leftmargin=29pt]
\item[1.] Load K, IV and constants into the LFSR for algorithm Initialization;
\item[2.] Perform KeystreamGeneration to produce a keystream of L=$\lceil l/32\rceil$  words. Denote the keystream Z by $\mathrm{Z_1\|Z_2\|\ldots\|Z_{\lceil \emph{l}/32 -1\rceil}\|
MSB_s(Z_{\lceil \emph{l}/32\rceil})}$, where Z$_i$ is the i-th output word and MSB$\mathrm{_s(Z_{\lceil \emph{l}/32\rceil})}$ is the most significant s-bit of $Z_{\lceil l/32\rceil}$, s = $l-32\times(\lceil l/32 -1\rceil)$.
\end{itemize}
\section{Specification of ZUC-GXM}\label{sec: 3}
\subsection{Design Framework}\label{subsec: 3.1}
Initially, GCM (Galois/Counter Mode of Operation) is a AEAD mode of operation based on blockciphers. It is designed by McGrew and Viega[2]. It is based on the counter mode encryption and the polynomial hash function GHASH for anthentication. GCM is the most widely used blockcpher modes. Following the framework of GCM, we give the  mode GXM for stream ciphers. The main difference lies in that GXM is no need to generate counters for each blocks, so we name it GXM fitly. From the prg property of stream ciphers, we can see its security bound for privacy is higher than GCM for block ciphers.
\subsection{Input, Output and Keys}\label{subsec: 3.2}
\textbf{Input.} There are three input strings to the authenticated encryption function:
\begin{itemize}
\item[]\hspace{10pt} $\scriptscriptstyle{\bullet}$ a plaintext, denoted P ;
\item[]\hspace{10pt} $\scriptscriptstyle{\bullet}$ additional associated data (AAD), denoted A ; and
\item[]\hspace{10pt} $\scriptscriptstyle{\bullet}$ an initialization vector (IV), denoted IV. The IV is essentially a nonce N, i.e, a value that is unique within the specified context.
\end{itemize}
\textbf{Output.} The output strings are:
\begin{itemize}
\item[]\hspace{10pt} $\scriptscriptstyle{\bullet}$ a ciphertext, denoted C ;
\item[]\hspace{10pt} $\scriptscriptstyle{\bullet}$ Authentication code, denoted Tag .
\end{itemize}
\textbf{Lengths.} The bit lengths of the input and output strings to the authentic-
cated encryption function shall meet the following requirements:
\begin{itemize}
\item[]\hspace{10pt} $\scriptscriptstyle{\bullet}$ $\mathrm{|Tag|=\tau, 32\leq\tau\leq128}$, we recommend to use 128 bits;

For 128-bit Tag, we requre:

\item[]\hspace{10pt} $\scriptscriptstyle{\bullet}$ $\mathrm{|P|+|A|+1< 2^{64}}$;
\item[]\hspace{10pt} $\scriptscriptstyle{\bullet}$ $\mathrm{|N|=v}$, v is determined by ZUC algorithm;
\item[]\hspace{10pt} $\scriptscriptstyle{\bullet}$ $\mathrm{|C|=|P|}$.
\end{itemize}
\textbf{Keys.} ZUC-GXM needs two keys, H and K. H is a 128 bits key for universal hash GHASH, and K is a k (128 or 256) bits secret key for ZUC encryption.
\subsection{Description of ZUC-GXM}\label{subsec: 3.3}
The ZUC-GXM authenticated encryption algorithm is given as follow.

\begin{figure}[htp]
\centering
\includegraphics[width=2.4in]{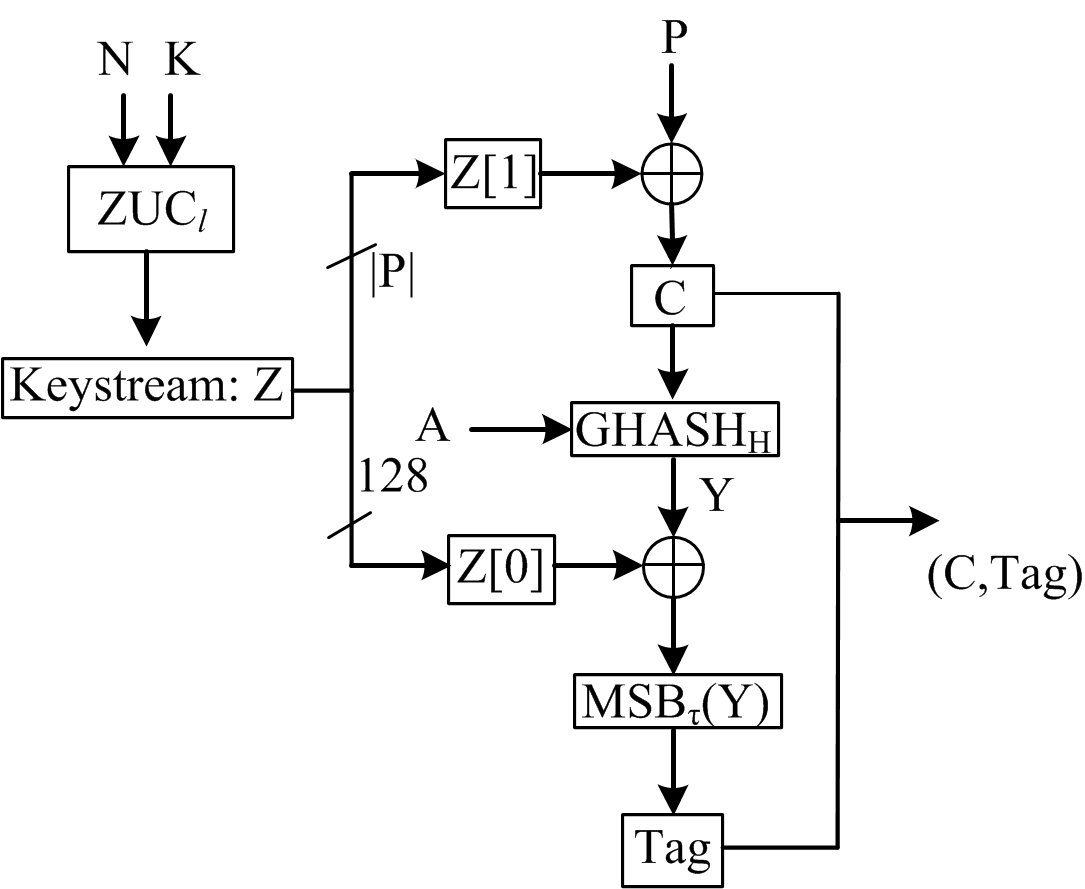}
\caption{ZUC-$\mathrm{GXM\_E}$}
\label{fig3.1}       
\end{figure}

\noindent\textbf{Algorithm} ZUC-$\mathrm{GXM_{H,K}(N, A,P)}$
\begin{itemize}
\item[1.] Z $\leftarrow$ ZUC$_l$(N, K), where $l$ = $\mathrm{|P|}$ + 128;
\item[2.] Let $\mathrm{Z=Z[0]\|Z[1]}$, where $Z[0]$ is of length 128 bits, $Z[1]$ is of length $\mathrm{|P|}$ bits;
\item[3.] $\mathrm{C \leftarrow P\oplus Z[1]}$;
\item[4.] $\mathrm{Y \leftarrow GHASH_H(A, C)}$, where $\mathrm{GHASH_H(\cdot)}$ is defined as follows;
\item[5.] $\mathrm{Tag \leftarrow MSB_\tau(Z[0]\oplus Y)}$;
\item[6.] Return (C, Tag).
\end{itemize}
\noindent\textbf{Algorithm} $\mathrm{GHASH_H(A, C)}$
\begin{itemize}
\item[1.]  Len$\leftarrow\mathrm{str_{n/2}(|A|)\parallel str_{n/2}(|C|)}$;
\item[2.]  X $\leftarrow\mathrm{A\|0^{n|A|_n-|A|}\parallel C \parallel 0^{n|P|_n-|P|}\parallel Len}$;
\item[3.]  $\mathrm{(X[1],\ldots, X[x])\xleftarrow{n}{}X}$;
\item[4.]  Y $\leftarrow 0^n$;
\item[5.]  For j = 1 to x do
\par5.1    \quad Y$\leftarrow\mathrm{H\cdot(Y\oplus X[j])}$, where $\cdot$ denotes multiplication in GF($2^{128}$) with polynomial p(x) (reference Section 2.1);
\item[6.]  Return Y.
\end{itemize}
The procedure is shown in Fig. \ref{fig3.1}.

\noindent The ZUC-GXM authenticated decryption algorithm is given bellow.

\noindent\textbf{Algorithm} ZUC-GXM$\mathrm{\_D_{H,K}(N, A, C, Tag)}$
\begin{itemize}
\item[1.] Z $\leftarrow$ ZUC$_\emph{l}$(N, \text{K}), where $l$ = $\mathrm{|C|}$ + 128;
\item[2.] Let $\mathrm{Z=Z[0]\|Z[1]}$, where $Z[0]$ is of length 128 bits, $Z[1]$ is of length $\mathrm{|C|}$ bits;
\item[3.] $\mathrm{Y\leftarrow GHASH_H(A, C)}$;
\item[4.] $\mathrm{Tag^\prime\leftarrow MSB_\tau(Z[0]\oplus GHASH_H(A,C))}$;
\item[5.] If $\mathrm{Tag^\prime\neq Tag}$, then return $\perp$;
\item[6.] $\mathrm{P\leftarrow C\oplus Z[1]}$;
\item[7.] Return P.
\end{itemize}

\begin{figure}[htp]
\centering
\includegraphics[width=2.4in]{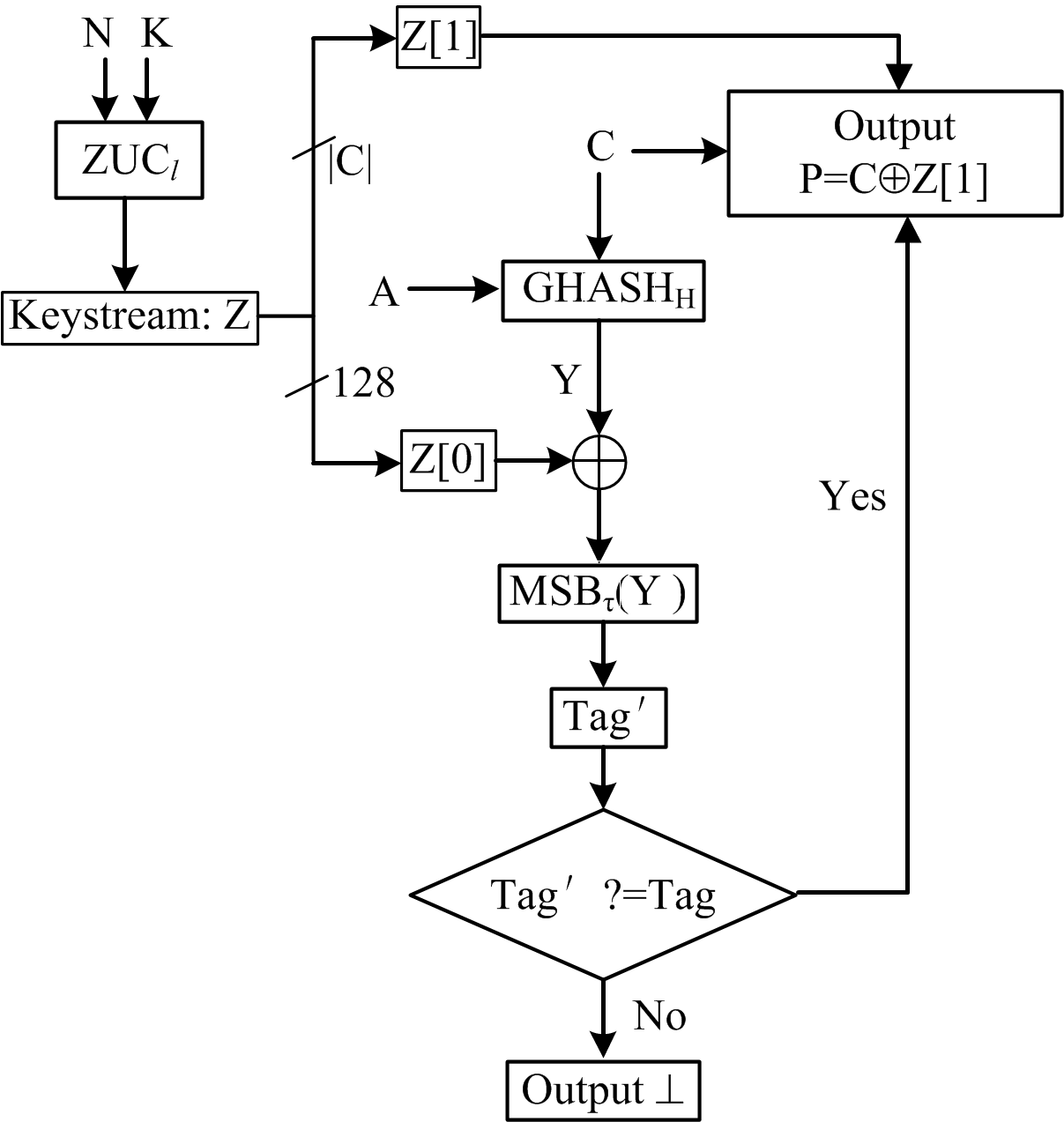}
\caption{ZUC-$\mathrm{GXM\_D}$}
\label{fig3.2}       
\end{figure}

Fig. \ref{fig3.2} illustrates the procedure.
\subsection{Proof Security of ZUC-GXM}\label{subsec: 3.4}
ZUC can be regard as a pseudorandom generator (prg). Further, We assume ZUC has the following prf property as described in section \ref{subsec:2.2}: for $\mathrm{K\in \widehat{K}}$, and $\mathrm{N_0, N_1\in\widehat{N}, N_1\neq N_0}$, then ZUC(N$_0$, K) and ZUC(N$_1$, K) are independent pseudorandom output.

Let $\mathrm{(N_1, A_1, P_1),\ldots,(N_q, A_q, P_q)}$ be all the encryption queries made by $\mathcal{A}$. The responses are $\mathrm{(N_1, A_1, C_1,}$  $\mathrm{T_1)}$ $\mathrm{\ldots,(N_q, A_q, C_q, T_q)}$ respec tively. And let $\mathrm{(N_1^\prime, A_1^\prime, C_1^\prime, T_1^\prime),\ldots,(N_q^\prime, A_q^\prime, C_q^\prime, T_{q_v}^\prime)}$ be all the decryption queries made by $\mathcal{A}$. For $\mathrm{1\leq i\leq q_v}$, outputs a plaintext $\mathrm{P_i^\prime}$ with $\mathrm{|P_i^\prime|=|C_i^\prime|}$ if input is determined as valid, or error symbol $\bot$, if determined as invalid.

We further assume $\mathcal{A}$ in the authenticity notion is always nonce-respecting with respect to encryption queries. Using the same N for encryption and decryption queries is allowed, and the same N can be repeated within decryption queries, i.e. $\mathrm{N_i}$ is different from $\mathrm{N_j}$ for any j $\neq$ i, but $\mathrm{N_{i^\prime}^\prime}$ may be equal to $\mathrm{N_{j^\prime}^\prime}$ or $\mathrm{N_i}$ for some i and i$^\prime\neq$ i.

Without loss of generality, we can assume $\mathcal{A}$ performs all encryption queries before the decryption query at first, which is the best strategy for maximizing the probability of successful forgery. We also assume that $\mathcal{A}$ don't perform the queries that their responses is already known, and he don't perform the same query more than once. Furthermore, we assume that $\mathcal{A}$ will abort when he forges. Then, for ZUC-GXM we have the following security results for a single user with once-used initial vector nonce.


\noindent\textbf{Theorem 1 (Privacy).} For any cpa-adversary $\mathcal{A}$ with parameter (t, q, $\sigma$), we have
\begin{equation*}
\mathrm{Adv_{ZUC\text{-}GXM}^{priv}(\mathcal{A})\leq Adv_{ZUC}^{prf}(t_1,q,\sigma)},
\end{equation*}
where q is the total number of encryption queries adversary $\mathcal{A}$ made, $\sigma$ is the total length in bits of those queries with running time t. And $\mathrm{t_1\approx t}$, and $\approx$ denotes the difference between its two sides is less than a polynomial of q and $\sigma$, which is depended on ZUC.

\noindent\textbf{(Authenticity).} For any cca-adversary $\mathcal{A}$ with parameter (t, q, $\mathrm{q_v, \sigma_1, \sigma_2)}$, we have
\begin{equation*}
\mathrm{Adv_{ZUC\text{-}GXM[\tau]}^{auth}(\mathcal{A})\leq Adv_{ZUC}^{prf}(t_2,q+q_v,\sigma)+q_v\varepsilon},
\end{equation*}
where $\mathrm{t_2\approx t, \sigma = \sigma_1 + \sigma_2 + 128(q + q_v)}$,
$\mathrm{\varepsilon = L/2^{\tau}}$, $\mathrm{L = L_0 + 1}$,

\noindent$\mathrm{L_{0} = max_{i\in\{1,\ldots,q + q_v\}}\left\{\left\lceil\frac{|P_i|}{128}\right\rceil + \left\lceil\frac{|A_i|}{128}\right\rceil\right\}.}$


\noindent\textbf{Proof.} Firstly, we give the privacy analysis, and then the authenticity proof is given.

\noindent\textbf{Privacy.} To prove the prf advantage $\mathrm{Adv_{ZUC\text{-}GXM}^{priv}}$ of ZUC-GXM, firstly we replace the output of ZUC by uniform random strings $\$$. And then we will give the distinguishing advantage between $\$$-GXM and uniform random output.
From the i-th encryption query $\mathrm{(N_i, A_i, P_i)}$ of $\mathcal{A}$. The result is a uniformly random string of length $\mathrm{|P_i|+128}$, independent of all other query responses. More specifically, for different N, $\$$ generates uniformly random keystream $\mathrm{Z_i}$, where $\mathrm{|Z_i|=|P_i|}$. Each $\mathrm{Z_i, 1\leq i\leq q}$ are independent of each other. $\mathrm{Z_i}$ is partitioned into two independent part, $\mathrm{Z_i[0]}$ and $\mathrm{Z_i[1]}$, and they are used for encryption and Tag generation respectively. And they are also independent with $\mathrm{Z_i[0]}$ and $\mathrm{Z_j[1]}$, for $\mathrm{j\neq i}$. Hence, the responding value $\mathrm{C_i\|T_i}$ of the i-th query is distinguishable from random strings of equal length.

From above analysis, we have
\begin{equation*}
\mathrm{Adv_{\$\text{-}GXM}^{priv}(\mathcal{A})=0},
\end{equation*}
Hence, for the privacy of ZUC-GXM, we have
\begin{align*}
&\mathrm{Adv_{ZUC\text{-}GXM}^{priv}(\mathcal{A})=Adv_{ZUC\text{-}GXM,\$}^{cpa}(\mathcal{A})}\\
\leq&\mathrm{Adv_{ZUC\text{-}GXM,\$\text{-}GXM}^{cpa}(\mathcal{A})+Adv_{\$\text{-}GXM,\$}^{cpa}(\mathcal{A})}\\
\leq&\mathrm{Adv_{ZUC,\$}^{cpa}(\mathcal{A}_1)+Adv_{\$\text{-}GXM}^{priv}(\mathcal{A}_2)}\\
\leq&\mathrm{Adv_{ZUC}^{prf}(t_1,q,\sigma)}.
\end{align*}

\noindent\textbf{Authenticity.} Similar to the privacy proof, we first give the authenticity of $\$$-GXM[$\tau$], by replacing the output of ZUC by uniform random strings $\$$. First, We give authenticity of $\$$-GXM[$\tau$] as follow.

For $\mathrm{1\leq i\leq q_v}$, suppose the real Tag of the query $\mathrm{(N_i^\prime,A_i^\prime,C_i^\prime,T_i^\prime)}$ is $\mathrm{T_i^\star}$. Then the authenticity proof is equivalent to analysis adversary's success probability to obtain $\mathrm{T_i^\prime}=\mathrm{T_i^\star}$ by encryption and decryption queries. From the definition, we have
\begin{equation*}
\mathrm{T_i^\star=MSB_\tau(Z_i^{\prime}[0]\oplus GHASH_H(A_i^\prime,C_i^\prime))}.
\end{equation*}
Following, we will discuss in two cases.

\noindent\textbf{Case 1.} If $\mathrm{N_i^\prime\notin\{N_1,\ldots,N_j\}}$, where $\mathrm{N_1,\ldots, N_j}$ denote the initial vectors in the query before the i-th decryption query, then the event $\mathrm{T_i^\prime}=\mathrm{T_i^\star}$ is equivalent to
\begin{equation*}
\mathrm{T_i^\prime=MSB_\tau(Z[0]\oplus GHASH_H(A_i^\prime,C_i^\prime))}.
\end{equation*}

From the discuss in privacy proof, we have known that Z[0] is a random string. So we have
\begin{equation*}
 \mathrm{Pr[T_i^\prime=T_i^\star]\leq1/2^\tau}.
\end{equation*}

\noindent\textbf{Case 2.} If $\mathrm{N_i^\prime\in\{N_1,\ldots,N_j\}}$, there exits some encryption query $\mathrm{(N_{j^\prime},A_{j^\prime},P_{j^\prime})}$ such that $\mathrm{N_i^\prime=N_{j^\prime}}$ . Let $\mathrm{T_{j^\prime}}$ and $\mathrm{C_{j^\prime}}$ be the authenticated code and the ciphertext respectively. Then $\mathrm{(A_i^\prime,C_i^\prime)\neq(A_{j^\prime},C_{j^\prime})}$ must be hold. And the event $\mathrm{T_i^\prime}=\mathrm{T_i^\star}$ is equivalent to
\begin{equation}
\mathrm{T_i^\prime\mspace{-3mu}\oplus \mspace{-3mu}T_{j^\prime}\mspace{-3mu}=\mspace{-3mu}MSB_\tau(Z_i^{\prime}[0]\mspace{-5mu}\oplus\mspace{-3mu} GHASH_H(A_i^\prime,C_i^\prime))\mspace{-3mu}\oplus\mspace{-3mu} Z_{j^\prime}[0]\mspace{-5mu}\oplus\mspace{-3mu} GHASH_H(A_{j^\prime}^\prime,C_{j^\prime}^\prime))\mspace{-3mu}=\mspace{-3mu}0}.\tag{*}\label{equ1}
\end{equation}
So, for any $\mathrm{Y\in\{0,1\}^n}$, the equation
\begin{equation*}
\mathrm{Y=GHASH_H(A_i^\prime,C_i^\prime)\oplus GHASH_H(A_{j^\prime},C_{j^\prime}))}
\end{equation*}
has at most L solutions, which implies that equation (\ref{equ1}) has at most $\mathrm{2^{n\text{-}\tau}L}$ solutions. Hence, for every $\mathrm{1\leq i\leq q_v}$ , we have
\begin{equation*}
\mathrm{Pr[T_i^\prime}=\mathrm{T_i^\star]\leq2^{n\text{-}\tau}L/2^n=L/2^\tau}.
\end{equation*}
Further we obtain,
\begin{equation*}
\mathrm{Adv_{\$\text{-}GXM[ZUC,\tau]}^{auth}(\mathcal{A})\leq q_v\varepsilon},
\end{equation*}

where $\mathrm{\varepsilon=L/2^\tau}$, $\mathrm{L = max_{i\in\{1,\ldots,q + q_v\}}\left\{\left\lceil\frac{|P_i|}{128}\right\rceil + \left\lceil\frac{|A_i|}{128}\right\rceil\right\} + 1}$.

From the above analysis, for the authenticity of ZUC-GXM[$\tau$], we have
\begin{align*}
\mathrm{Adv_{ZUC\text{-}GXM[\tau]}^{auth}(\mathcal{A})}&\leq \mathrm{Adv_{ZUC\text{-}GXM[\tau],\$\text{-}GXM[\tau]}^{cca}(\mathcal{A})+Adv_{\$\text{-}GXM[\tau]}^{auth}(\mathcal{A})}\\
&\leq\mathrm{Adv_{ZUC}^{prf}(t_2,q+q_v,\sigma)+q_v\varepsilon},
\end{align*}
where $\mathrm{t_2\approx t, \sigma = \sigma_1 + \sigma_2 + 128(q + q_v)}.\qquad\qquad\qquad\square$
\section{Specification of ZUC-MUR}\label{sec: 4}
\subsection{Design Framework}\label{subsec: 4.1}
\begin{figure}[htp]
\centering
\includegraphics[width=1.5in]{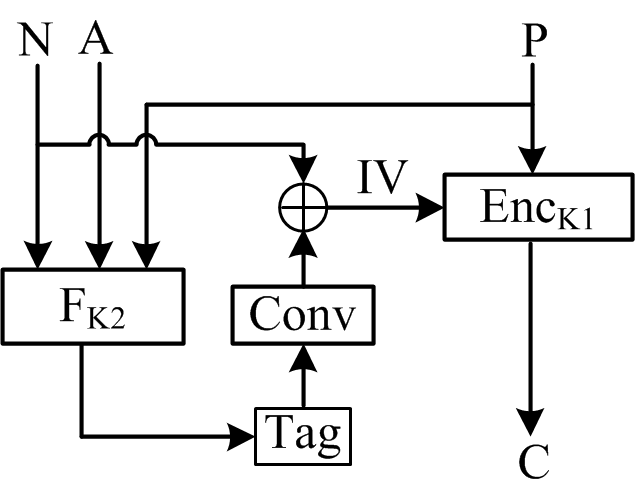}
\caption{MUR Design Framework}
\label{fig4.1}       
\end{figure}

\begin{figure}[htp]
\centering
\includegraphics[width=1.5in]{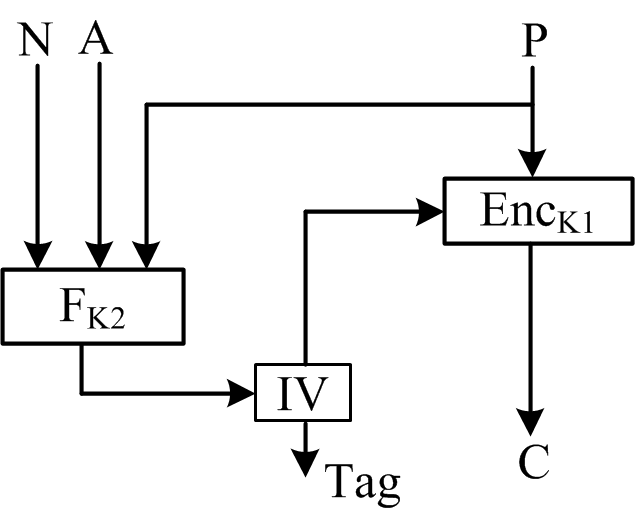}
\caption{SIV}
\label{fig4.2}       
\end{figure}

We first give an variance of SIV framework (see Fig. \ref{fig4.2}), and then design a misuse-resistant AEAD mode MUR based on it. Unlike the SIV framework, in MUR, the nonce N participates not only in the Tag generation phase, but also in the encryption phase as depicted in Fig. \ref{fig4.1}. By doing so, the security bound won't be decreased when the N is used as normal. See section \ref{sec: 5} for the details.

\begin{figure}[htp]
\centering
\includegraphics[width=1.8in]{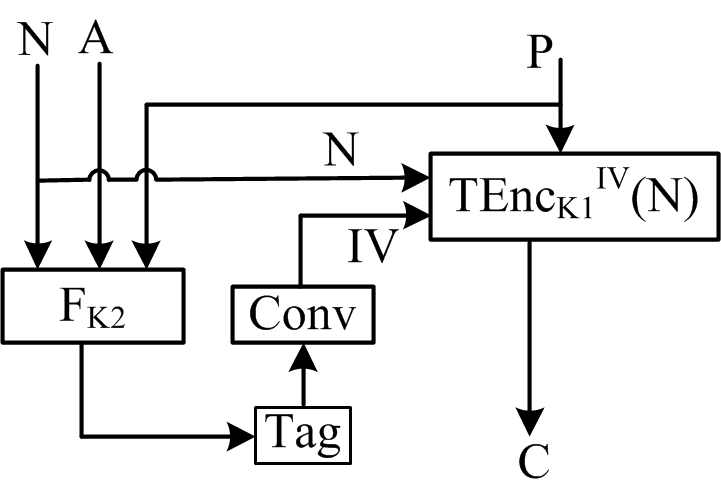}
\caption{NSIV}
\label{fig4.3}       
\end{figure}

Besides SIV, very similar construction is NSIV (see Fig. \ref{fig4.3}). But it is different from ours in how to use N and Tag in the encryption function. That is, for NSIV, the Tag (IV) is used as a tweak for encryption, while in our scheme, the XORed value of N and Tag is used as the initial vector of the encryption function. The main differeces among the three high level constructions are shown in Fig. \ref{fig4.1} - Fig. \ref{fig4.3}.
\subsection{Specification of ZUC-MUR}\label{subsec: 4.2}
\subsubsection{Input,Output and Keys}
The input, output and their length requirements are the same as in ZUC-GXM. And ZUC-MUR also needs two keys, H and K (we use K$_1$=K$_2$=K), where H is a 128 bits key for universal hash GHASH, and K is a k (128 or 256) bits secret key for ZUC for both authentication and encryption.
\subsubsection{ZUC-MUR description}
The ZUC-MUR authenticated encryption algorithm is given as follows.

\noindent\textbf{Algorithm} ZUC-MUR$\mathrm{\_E_{H,K}(N, A, P)}$
\begin{itemize}
\item[1.] Y$\leftarrow$GHASH$_\text{H}$(A, P), same as defined in section \ref{subsec: 3.3};
\item[2.] $\mathrm{Tag\leftarrow ZUC_\tau(Conv(Y)\oplus N, K)}$;
\item[3.] $\mathrm{Z\leftarrow ZUC_{|P|}(Conv(Tag)\oplus N, K)}$;
\item[4.] $\mathrm{C\leftarrow P\oplus Z}$;
\item[5.] Return (C, Tag);
\end{itemize}
where Conv($\cdot$) is a regular function defined bellow:
\begin{equation*}
\mathrm{y=Conv(Y)=}\left\{
\begin{aligned}
&\mathrm{Y,\qquad\qquad if\ |Y|=v;}\\
&\mathrm{Y\|0^{v-|Y|},\quad else\ if\ |Y|<v.}\\
\end{aligned}
\right.
\end{equation*}

\begin{figure}[htp]
\centering
\includegraphics[width=2.0in]{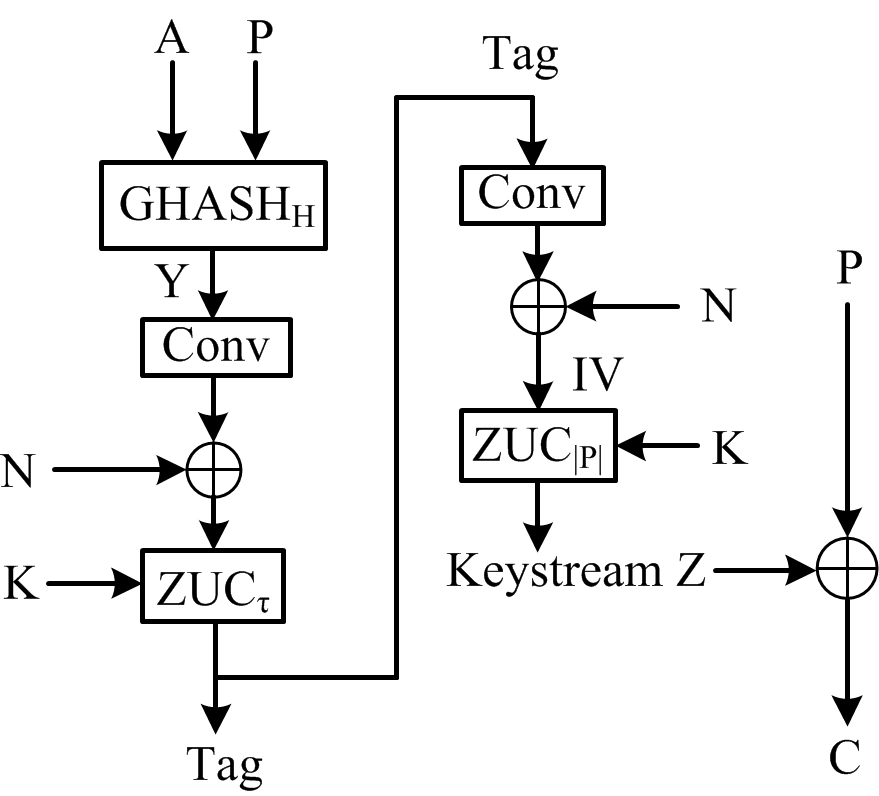}
\caption{ZUC-MUR$\_$E}
\label{fig4.4}       
\end{figure}

Fig. \ref{fig4.4} illustrates the procedure.

The following is the ZUC-MUR$\_$D verification and decryption algorithm.

\noindent\textbf{Algorithm} ZUC-MUR$\mathrm{\_D_{H,K}(N, A, C, Tag)}$
\begin{itemize}
\item[1.] $\mathrm{Z\leftarrow ZUC_{|C|}(Conv(Tag)\oplus N, K)}$;
\item[2.] $\mathrm{P\leftarrow C\oplus Z}$;
\item[3.] $\mathrm{Y\leftarrow GHASH_H(A,P)}$;
\item[4.] $\mathrm{Tag^\prime\leftarrow ZUC_{\tau}(Conv(Y)\oplus N, K)}$;
\item[5.] If $\mathrm{Tag^\prime\neq Tag}$, delete the decrypted plaintext and return $\perp$;
\item[6.] Return P.
\end{itemize}

\begin{figure}[htp]
\centering
\includegraphics[width=3.5in]{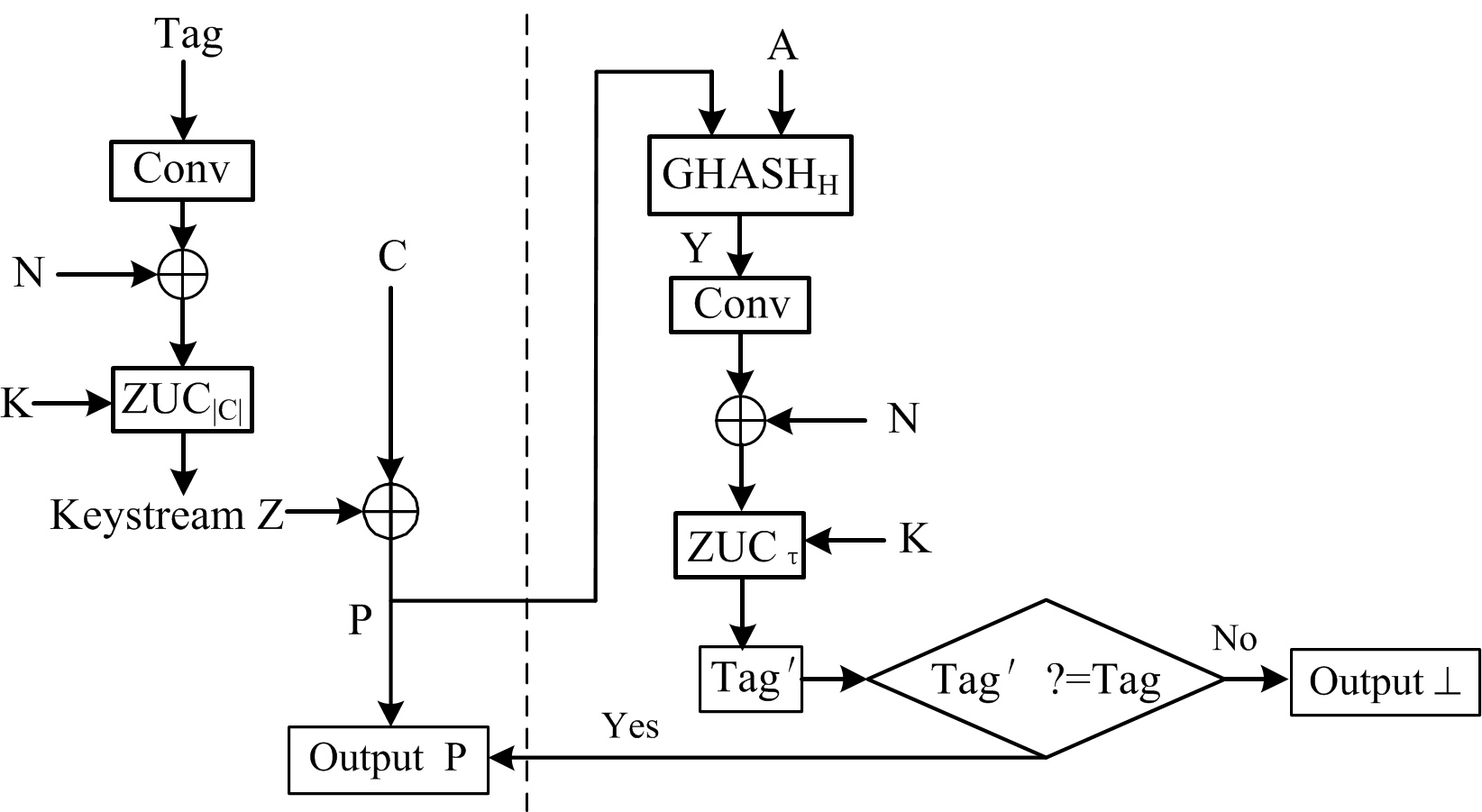}
\caption{ZUC-MUR$\_$D}
\label{fig4.5}       
\end{figure}

The algorithm procedure is shown in Fig.\ref{fig4.5}.
\subsection{Proof Security of ZUC-MUR}\label{subsec: 4.3}
\subsubsection{Nonce-misuse security of ZUC-MUR}
For the security of ZUC-MUR , we have the main result as given below.

\noindent\textbf{Theorem 2.} Let $\mathcal{A}$ be any cca-adversary with parameters ($\mathrm{q_E, q_D, \sigma_E, \sigma_D}$) with running time t, where $\mathrm{q_E}$ is the encryption queries he made and $\mathrm{\sigma_E}$ is the total length of the plaintexts in bits of the encryption queries, $\mathrm{q_D}$ and $\mathrm{\sigma_D}$ means similarly. Then, for the misuse-resistant security of ZUC-MUR, we have
\begin{align*}
\mathrm{Adv_{ZUC\text{-}MUR[\tau]}^{mrAEAD}(\mathcal{A})}&\leq \mathrm{Adv_{ZUC}^{prf}(t,q_f,q_f\tau)+Adv_{ZUC}^{prf}(t,q_f,\sigma)}\\
&+\mathrm{\varepsilon_1\cdot\binom{q_f}{2}+\varepsilon_2\cdot\binom{q_E}{2}+q_D/2^\tau},
\end{align*}
where $\mathrm{q_f=q_E+q_D, \sigma=\sigma_E+\sigma_D, \varepsilon_1=L/2^{128}, \varepsilon_2=1/2^\tau}$, and
\begin{equation*}
\mathrm{L = max_{i\in\{1,\ldots,q + q_v\}}\left\{\left\lceil\frac{|P_i|}{128}\right\rceil + \left\lceil\frac{|A_i|}{128}\right\rceil\right\} + 1},
\end{equation*}
where $\mathrm{P_i}$ and $\mathrm{A_i}$ are respectively the paintext and the associated data in the i-th query.

\begin{figure}[htp]
\centering
\includegraphics[width=1.1in]{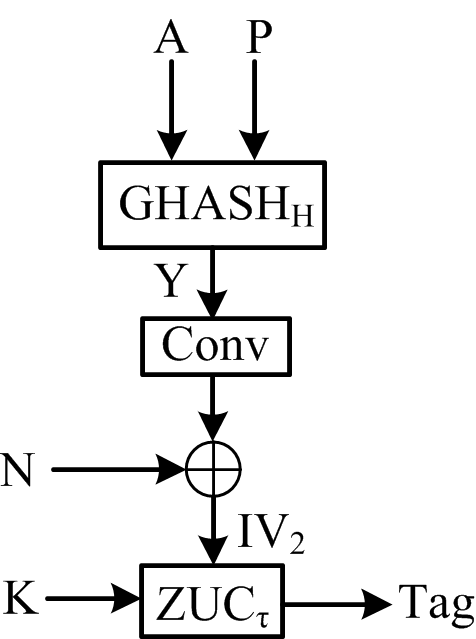}
\caption{TagG. scheme}
\label{fig4.6}       
\end{figure}

\begin{figure}[htp]
\centering
\includegraphics[width=1.2in]{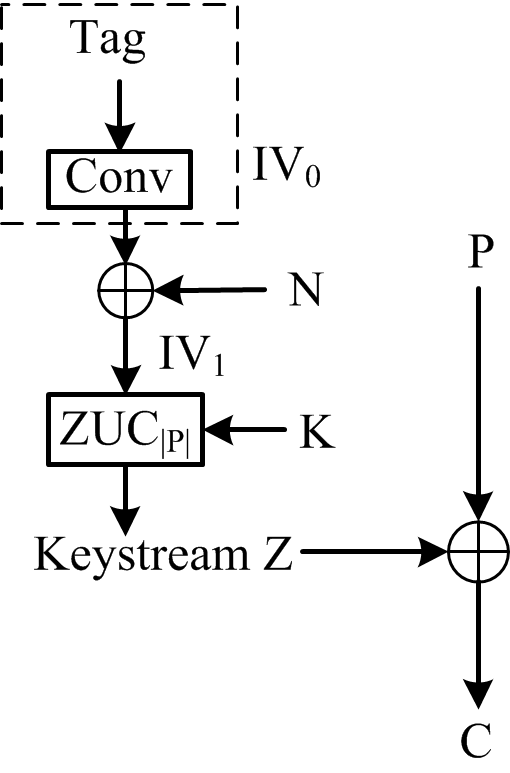}
\caption{Enc. scheme}
\label{fig4.7}       
\end{figure}

According to section \ref{sec: 4}, ZUC-MUR authenticated encryption procedure consists of two parts: the Authenti-cation-code-generation scheme \textbf{TagG.} (Fig.\ref{fig4.6}) and the encryption scheme \textbf{Enc.} (Fig.\ref{fig4.7}).

We define
\begin{equation}
\mathrm{TagG_{H,K}:=ZUC_\tau(IV_2,K)},\label{equ4.1}
\end{equation}
where $\mathrm{Y=GHASH_H(A, P), IV_2= Conv(Y)\oplus N}$.
\begin{equation}
\mathrm{Enc_{K_1}:=ZUC_{|P|}[IV_1,K]\oplus P, where\ IV_1 = IV_0\oplus N}.\label{equ4.2}
\end{equation}

In our scheme, we use K$_1$=K$_2$=K because it can be ensured that the probability of IV$_1$=IV$_2$ is negligeable. So, similar to the security of SIV\cite{8} and NSIV\cite{5}, we can easily obtain the following lemma.

\noindent\textbf{Lemma 1.} Let $\mathcal{A}$ be a ($\mathrm{q_f, q_E, q_D, B}$)-adversary, there exist adversaries $\mathcal{A}_1$ and $\mathcal{A}_2$ with $\mathcal{A} = \mathcal{A}_1+ \mathcal{A}_2$, where $\mathcal{A}_1$ makes at most $\mathrm{q_f}$ TagG function queries, $\mathcal{A}_2$ makes at most $\mathrm{q_E}$ encryption queries and $\mathrm{q_D}$ decryption queries, and the total length of message sent by $\mathcal{A}$ to the oracles is B, then
\begin{equation*}
\mathrm{Adv_{ZUC\text{-}MUR}^{mrAEAD}(\mathcal{A})\leq Adv_{TagG}^{prf}(\mathcal{A}_1)+Adv_{Enc}^{priv}(\mathcal{A}_2)+q_D/2^\tau}.
\end{equation*}

From the above Lemma, in order to prove the security of ZUC-MUR, we only need to condider the prf-advantage of function TagG, and the privacy of function $\mathrm{Enc_{K_2}}$ respectively.

For the prf-advantage of function TagG , we assume that Tag = $\mathrm{ZUC_\tau(IV, K)}$ is pseudorandom function restricted in $\mathrm{\{0, 1\}^v\times\{0, 1\}^{256}\rightarrow\{0, 1\}^\tau}$. Then we have the following lemma hold.

\noindent\textbf{Lemma 2.} Let adversary $\mathcal{A}_1$ make $\mathrm{q_f}$ times queries with (N, A, P). Then equation (\ref{equ4.1}) is a pseudorandom function about (N, A, P), that is
\begin{equation*}
\mathrm{Adv_{TagG}^{prf}(\mathcal{A}_1)\leq Adv_{ZUC}^{prf}(t,q_f,q_f\tau)+\varepsilon_1\cdot\binom{q_f}{2}},
\end{equation*}
where $\mathrm{q_f=q_E+q_D, \varepsilon_1=L/2^{128}}$, and
\begin{equation*}
\mathrm{L = max_{i\in\{1,\ldots,q + q_v\}}\left\{\left\lceil\frac{|P_i|}{128}\right\rceil + \left\lceil\frac{|A_i|}{128}\right\rceil\right\} + 1}.
\end{equation*}
where $\mathrm{P_i}$ and $\mathrm{A_i}$ are respectively the paintext and the associated data in the i-th qury.

\noindent\textbf{Proof.} Firstly, according to the description of ZUC-MUR, we give the following definitions. Let
\begin{align*}
&\mathrm{W_0=TagG_{H,K}:=ZUC_\tau\{Con[GHAH_H(A,P)]\oplus N, K\}}, \text{where }\\
&\qquad\quad\text{ H and K are random-chosen in the key spaces. }\\
&\mathrm{W_1=R_\tau\{Con[GHASH_H(A,P)]\oplus N,K\}, \text{where R is a uniform}}\\
&\qquad\quad\text{ random function, and }\\
&\mathrm{W_2=\$_\tau(N,A,P)}.
\end{align*}
Then we have
\begin{align*}
\mathrm{Adv_{TagG}^{prf}(\mathcal{A}_1)}&\mathrm{=|Pr[\mathcal{A}^{w_0}\Longrightarrow1]-Pr[\mathcal{A}^{w_1}\Longrightarrow1]|}+\\
&+\mathrm{|Pr[\mathcal{A}^{w_1}\Longrightarrow1]-Pr[\mathcal{A}^{w_2}\Longrightarrow1]|}\\
&\leq\mathrm{Adv_{ZUC}^{prf}(\mathcal{A}_1)+p},
\end{align*}
where $\mathrm{p = |Pr[\mathcal{A}^{w_1}\Longrightarrow1]-Pr[\mathcal{A}^{w_2}\Longrightarrow1]|}$. We need only estimate the remaining probability p.

From GCM, we know that the GHASH in W$_1$ have $\mathrm{\varepsilon\text{-}AXU}$ property, where $\mathrm{=L/2^\tau}$, L is the maximum 128-bit blocks of ternary ($\mathrm{N_i, A_i, P_i}$).

Let (N$_1$,A$_1$,P$_1$),$\mathrm{\ldots,(N_{q_f},A_{q_f},P_{q_f})}$ be the q$_f$ different queries. And let

$\mathrm{Y_i= GHASH_H(A_i, P_i), Y_j= GHASH_H(A_j, P_j)}$. By the $\mathrm{\varepsilon\text{-}AXU}$ property of GHASH, we obtain
\begin{align*}
\text{p}&=\mathrm{Pr[\exists i, j\in\{1,2,\ldots,q_f\}|Y_i\oplus Y_j = N_i\oplus N_j]}\\
&\leq \mathrm{\sum_{i=1}^{q_f-1}\sum_{j=i+1}^{q_f}Pr[Y_i\oplus Y_j = N_i\oplus N_j]}\\
&\leq\mathrm{\varepsilon_1\cdot \binom{q_f}{2}}.
\end{align*}
Hence, by equation (\ref{equ4.1}), we have TagG' prf-advantage is
\begin{equation*}
\mathrm{Adv_{TagG}^{prf}(\mathcal{A}_1)\leq Adv_{ZUC}^{prf}(t,q_f,q_f\tau)+\varepsilon_1\cdot\binom{q_f}{2}}.\qquad\square
\end{equation*}

For the privacy of encryption scheme $\mathrm{Enc_{K_2}}$, we have the following lemma.

\noindent\textbf{Lemma 3.} Let adversary $\mathcal{A}_2$ make at most $\mathrm{q_E}$ encryption queries (N, A, P), then the equation (\ref{equ4.2}) is a pseudorandom function about (N, A, P), that is
\begin{equation*}
\mathrm{Adv_{Enc}^{priv}(\mathcal{A}_2)\leq Adv_{ZUC}^{prf}(t,q_E,\sigma_E)+\varepsilon_2\cdot\binom{q_E}{2}},
\end{equation*}
where $\mathrm{\sigma_E}$ is the total length of the plaintext in bits of his encryption queries, and $\mathrm{\varepsilon_2 = 1/2^\tau}$.

\noindent\textbf{Proof.} Obviously, as long as the collision do not happen between two queries, the ZUC in equation (\ref{equ4.2}) will generate undistinguishable keystreams. So we can obtain the ciphertext is also undistinguishable from random.

For $\mathrm{q_E}$ encryption queries $\mathrm{(N_i, A_i, P_i), i=1,2,\ldots,q_E}$, the collision probability of tags is as follows.
\begin{align*}
\text{p}&=\mathrm{Pr[\exists i, j\in\{1,2,\ldots,q_E\}|Tag_i\oplus Tag_j=0]}\\
&\leq \mathrm{\Sigma_{i=1}^{q_E-1}\Sigma_{j=i+1}^{q_E}Pr[Tag_i\oplus Tag_j=0]}\\
&\leq\mathrm{\varepsilon_2\cdot \binom{q_E}{2},\ where\ \varepsilon_2 = 1/2^\tau}.
\end{align*}
Hence, we have
\begin{equation*}
\mathrm{Adv_{Enc}^{priv}(\mathcal{A}_2)\leq Adv_{ZUC}^{prf}(t,q_E,\sigma_E)+\varepsilon_2\cdot\binom{q_E}{2}}.\qquad\square
\end{equation*}

\noindent\textbf{Proof of Theorem 2.} For each encryption query, the maximum 128-bit
-block length of the ternary ($\mathrm{N_i,A_i,P_i}$) is L, that is $\mathrm{|N_i|_{128}+|A_i|_{128}+|P_i|_{128}\leq L}$.
For ($\mathrm{N_i,A_i,C_i}$), we also have $\mathrm{|N_i|_{128}+|A_i|_{128}+|P_i|_{128}\leq L}$.

So, by Lemma 1 to Lemma 3, we obtain
\begin{align*}
\mathrm{Adv_{ZUC\text{-}MUR[\tau]}^{mrAEAD}(\mathcal{A})}&\leq \mathrm{Adv_{ZUC}^{prf}(t,q_f,q_f\tau)+Adv_{ZUC}^{prf}(t,q_f,\sigma)}\\
&+\mathrm{\varepsilon_1\cdot\binom{q_f}{2}+\varepsilon_2\cdot\binom{q_E}{2}+q_D/2^\tau},
\end{align*}
where $\mathrm{q_f=q_E+q_D, \sigma=\sigma_E+\sigma_D, \varepsilon_1=L/2^{128}, \varepsilon_2=1/2^\tau}$. \qquad $\square$
\subsubsection{Nonce-respect Security of ZUC-MUR}
For the privacy and authentication of ZUC-MUR in nonce-respect setting, if we make some restriction on the initial vector N, then we have the following theorem.

\noindent\textbf{Theorem 3 (privacy).} For any cpa-adversary $\mathcal{A}$ with parameter (t, q, $\sigma$), where q is the total number of encryption queries adversary $\mathcal{A}$ made, $\sigma$ is the total length of the plaintext in bits of those queries with running time t. Then we have
\begin{itemize}
\item[(a)] If the least significant $v-\tau$ bits of N is not reused, we have
\begin{equation*}
\mathrm{Adv_{ZUC\text{-}MUR[\tau]}^{priv}(\mathcal{A})\leq Adv_{ZUC}^{prf}(t_1,q_E,\sigma)}.
\end{equation*}
where $\mathrm{t_1\approx t}.$
\item[(b)] If N is random-generated, we have
\begin{equation*}
\mathrm{Adv_{ZUC\text{-}MUR[\tau]}^{priv}(\mathcal{A})\leq Adv_{ZUC}^{prf}(t,q_E,\sigma)+\varepsilon\cdot\binom{q_E}{2},}
\end{equation*}
\end{itemize}
 where $\mathrm{\varepsilon=1/2^v}$.

\noindent\textbf{(Authenticity).} For any ($\mathrm{q_E, q_D,\sigma_E,\sigma_D}$)-adversary $\mathcal{A}$ with running time t,
Then we have
\begin{equation*}
\mathrm{Adv_{ZUC-MUR[\tau]}^{auth}(\mathcal{A})\leq Adv_{ZUC}^{prf}(t,q_f,\sigma)+Adv_{ZUC}^{prf}(t,q_f,q_f\tau)+q_D\varepsilon},
\end{equation*}
where $\sigma_E$ is the total length of the plaintext in bits of the encryption queries, and $\sigma_D$ means similarly. $\mathrm{\sigma = \sigma_E+\sigma_D, q_f = q_E + q_D, \varepsilon = L/2^{\tau}}$, and
\begin{equation*}
\mathrm{L = max_{i\in\{1,\ldots,q + q_v\}}\left\{\left\lceil\frac{|P_i|}{128}\right\rceil + \left\lceil\frac{|A_i|}{128}\right\rceil\right\} + 1},
\end{equation*}
where $\mathrm{P_i}$ and $\mathrm{A_i}$ are respectively the paintext and the associated data in the i-th query.

\noindent\textbf{Proof.} Since the proof is similar to Theorem 1, we ignor it here. $\square$

From the above theorem, we can see when ZUC-MUR is used as a nonce-rspect AEAD, the security bound of ZUC-MUR will be preserved, when the lower $v-\tau$ bits of N is not reused.

\section{Reducing the Number of Keys}\label{sec: 5}
ZUC-GXM and ZUC-MUR need two independent keys. In order to reduce the storagement, as a alternative, we give a key derivation scheme for them without effecting their security.

\noindent\textbf{DeriveKey.} Let IV$_0$ be the initial vector of length v-bit, which is a system parameter and K$_0$ be the key of length k bits. Then the key derivation function DeriveKey(IV$_0$, K$_0$) can generate a 128-bit GHASH key H and a k-bit key K by performing ZUC. Here the value v and k is determined by which algorithm in the ZUC family is used. Then DeriveKey is defined as:

\qquad\qquad\qquad\qquad H$\parallel$K = ZUC$_{128+k}$(IV$_0$, K$_0$).
%

DeriveKey is designed as a spare scheme for reducing system storagement.
\section{Conclusion}\label{sec: 6}
We have presented two AEAD schemes for ZUC, ZUC-GXM and ZUC-MUR. ZUC-GXM is nonce-respecting, that is, it requires the initial vector not be reused for different message under the same secret key. And ZUC-MUR is a nonce-misuse resistant. It can guarantee that if a nonce is reused then the only damage is that an adversary can know if the associated data and plaintext message in two messages with the same nonce are the same or different, but nothing more. Therefore, ZUC-MUR can provide more robust applications, while the former is more efficient. Finally, we also give a DeriveKey mechanisism for the two schemes.

\textbf{Comparing ZUC-GXM with SNOW-V-AEAD.} In their design paper \cite{7}, the authors also give SNOW-V-AEAD mode following the GCM design framework. The difference lie in that its secret key of GHASH is changed with each message. This is not convenient to improve the efficiency of GHASH by precomputing the multiplication tables, which may lead to efficiency problem. Differently, we use fix secret key for GHASH in ZUC-GXM, in order to improve the efficiency conveniently.



\end{document}